%% file: bare_jrnl.tex
\documentclass[journal]{IEEEtran}
\usepackage{color}
\usepackage{array}
\usepackage{algorithm}
\usepackage{amsmath}
\usepackage{algorithmicx}
\usepackage{algpseudocode}
\usepackage{cite}
\usepackage{lettrine}

\usepackage{amssymb}

\usepackage{graphicx}

\ifCLASSINFOpdf
\else
\fi

\begin{document}

\title{Joint Signal Detection and Automatic Modulation Classification via Deep Learning}

\author{Huijun Xing,~\IEEEmembership{Student Member,~IEEE,}
        Xuhui Zhang,~\IEEEmembership{Student Member,~IEEE,}
        Shuo Chang,~\IEEEmembership{Member,~IEEE,}\\
        Jinke Ren,~\IEEEmembership{Member,~IEEE,}
        Zixun Zhang,~\IEEEmembership{Student Member,~IEEE,}
        Jie Xu,~\IEEEmembership{Senior Member,~IEEE,}\\
        and Shuguang Cui,~\IEEEmembership{Fellow,~IEEE}


\thanks{
H. Xing, X. Zhang, J. Ren and Z. Zhang are with the Future Network of Intelligence Institute (FNii), the School of Science and Engineering (SSE), and the Guangdong Provincial Key Laboratory of Future Networks of Intelligence, The Chinese University of Hong Kong, Shenzhen, Guangdong 518172, China (e-mail: huijunxing@link.cuhk.edu.cn; xu.hui.zhang@foxmail.com; 
jinkeren@cuhk.edu.cn;
zixunzhang@link.cuhk.edu.cn).
}
\thanks{
S. Chang is with the School of Cyberspace Security, Beijing University of Posts and Telecommunications, Beijing 100876, China (e-mail: changshuo@bupt.edu.cn).
}
\thanks{
J. Xu and S. Cui are with the SSE, the FNii, and the Guangdong Provincial Key Laboratory of Future Networks of Intelligence, The Chinese University of Hong Kong, Shenzhen, Guangdong 518172, China (e-mail: xujie@cuhk.edu.cn;
e-mail: shuguangcui@cuhk.edu.cn).
}

\thanks{
\textit{(Huijun Xing and Xuhui Zhang contributed
equally to this work)}
\textit{(Corresponding author: Shuguang Cui)}
}

\thanks{This work is supported in part by National Natural Science Foundation of China under Grant (62201090)}
}

\maketitle

\begin{abstract}
Signal detection and modulation classification are two crucial tasks in various wireless communication systems. Different from prior works that investigate them independently, this paper studies the joint signal detection and automatic modulation classification (AMC) by considering a realistic and complex scenario, in which multiple signals with different modulation schemes coexist at different carrier frequencies. We first generate a coexisting RADIOML dataset (CRML23) to facilitate the joint design. Different from the publicly available AMC dataset ignoring the signal detection step and containing only one signal, our synthetic dataset covers the more realistic multiple-signal coexisting scenario.
Then, we present a joint framework for detection and classification (JDM) for such a multiple-signal coexisting environment, which consists of two modules for signal detection and AMC, respectively. In particular, these two modules are interconnected using a designated data structure called ``proposal". 
Finally, we conduct extensive simulations over the newly developed dataset, which demonstrate the effectiveness of our designs.
Our code and dataset are now available as open-source resources\footnote{https://github.com/Singingkettle/ChangShuoRadioData}.
\end{abstract}


\IEEEpeerreviewmaketitle

\section{Introduction}
The recent advancements in cognitive radio allow more efficient utilization of scarce spectrum resources and enable more flexible wireless communications. However, cognitive radio also introduces new technical challenges on signal detection and modulation classification,
which are becoming vital topics in both academia and industry
\cite{9454255}. Effective signal detection and modulation classification can help provide flexible spectrum management \cite{8950109}, identify co-channel interference \cite{9667215}, and ensure physical-layer security against various threats, such as pilot jamming and deceptive jamming  \cite{9462447}.

Signal detection, which is also known as spectrum sensing, is an important technique to detect whether a certain user (e.g., a primary user in cognitive radio systems \cite{10.1145/1409944.1409948}) is transmitting signals. Conventionally, signal detection is implemented based on energy detection and feature detection methods \cite{10.1145/1409944.1409948}. In particular, energy detection is performed by comparing the received signal power level with an appropriate decision threshold. This method is simple in implementation and can be accomplished within a short sensing time. However, its performance varies sharply as the noise power may change over time. In contrast, the feature detection method utilizes specific signal signatures, such as pilot, field sync, segment sync, or cyclostationarity, to detect signals \cite{10.1145/1234388.1234391}, which improves the efficiency and robustness to the noise, while costs of higher implementation complexity to extract features from raw signal data, with a possibility of requirement on prior knowledge.

On the other hand, 
automatic modulation classification (AMC) has become a critical technology in modern communication systems \cite{9764618}. In general, AMC methods can be classified into two categories, namely the likelihood-based (LB) methods and the feature extraction and representation-based (FB) methods, respectively. The LB methods construct multiple hypothesis testing problems to classify modulations, which may induce significant computational overhead and are generally challenging to be implemented in practice \cite{5606206}. In contrast, the FB methods classify modulations by extracting essential signal features \cite{dobre2007survey}, which can thus significantly reduce the computational complexity without sacrificing the classification accuracy.

In recent years, deep learning (DL) has achieved astonishing success in object detection and classification \cite{8114708}. In view of its great potential, the exploitation of deep learning for signal detection and AMC has received increasing attention from both academia and industry.
Specifically, the DL-based approaches can learn the underlying features within the signal data, thus enabling full exploration of the intrinsic data connections.
In general, the implementation of DL usually requires a large amount of data to train a neural network model, which can be collected in practical communication systems \cite{8418751}.
Different from conventional feature extraction methods that require proper feature selection and thus cannot adapt to time-varying channel environments, the DL-based methods do not need the feature extraction procedure and thus are suitable for different environments \cite{9462447}.

The deployment of DL in AMC highly depends on the quality of the training dataset. To this end, a variety of datasets have been developed, including RADIOML.2016.10A (RML16) \cite{o2016convolutional}, RADIOML.2018.01A (RML18) \cite{o2018over}, and RML22 \cite{sathyanarayanan2023rml22}. 
However, these datasets have the following two limitations. First, the existing datasets are only applicable for specific scenarios with limited generalization. For instance, the RML22 can only be utilized to recognize a single signal within a particular frequency. Therefore, developing a general dataset with multiple signals at different frequencies is of paramount importance for the development of DL-based methods for modulation classification.
Next, current AMC datasets often omit the detection process, by ignoring the fact that the signal detection task needs to be performed prior to AMC in practice. This thus leads to an overly idealized representation, which fails to capture the complexities and challenges of real-world modulation recognition tasks. As a result, it is necessary to design datasets capturing the effect of signal detection prior to AMC. These issues thus motivate our investigation in this work.

\subsection{Contributions}
To overcome the above limitations, this paper proposes a new DL-based method for joint signal detection and AMC. We first generate a simulated dataset, namely the coexisting RADIOML dataset (CRML23), to achieve better data performance with lower overhead. In particular, synthetic data generation exploits the mature signal and channel models in wireless communications to achieve a high degree of fit with real data. As compared with the existing datasets, our proposed dataset CRML23 incorporates multiple signals coexisting in a specific interval. These signals are randomly distributed within the specified range and exhibit distinct features, thus simulating the signal distribution in real-world scenarios.
Based on CRML23, we propose a novel DL-based joint framework for detection and classification (JDM), and assess its performance via extensive experiments. The main contributions of this paper are summarized as follows.
\begin{enumerate}[]
\item We propose a new framework, JDM, that can simultaneously achieve signal detection and modulation classification. JDM consists of two interconnected modules for signal detection and modulation classification, respectively. These two modules sequentially process the source data and establish internal connections by using a designated data structure called ``proposal". In particular, the training results and ground-truth data simultaneously influence both modules, fostering an optimized learning process from diverse feature perspectives. 

\item In the proposed JDM, the detection module is realized by a multiple-layer convolutional neural network (CNN) for extracting signal features and predicting the center frequency and bandwidth of the signal.
Furthermore, a band prediction model named proposal is incorporated at the end of the detection module to generate predictions regarding the frequency location of the signal, which also connects the following classification stage.
In addition, we adopt a data representation to output predictions, thereby exploiting the intrinsic relationships within the original data.

\item We conduct extensive experiments to evaluate the performance of JDM and show the impact of different parameters, such as signal-to-noise ratio (SNR), channel characteristics, Doppler effect, and clock offset, on the performance. Simulation results show that our proposed framework achieves higher accuracy in detection and classification than conventional approaches. 
\end{enumerate}

\subsection{Organizations}
The remainder of this paper is organized as follows. Section II introduces the related works on signal detection and AMC. Section III describes the signal model and the design objective. Section IV presents the generation of dataset CRML23. Section V proposes the JDM. Section VI presents experiment results to show the performance of the proposed framework. Finally, Section VII concludes this paper.

\section{Related Works}
\subsection{Signal Detection}
There have been various prior works on conventional signal detection designs \cite{10.1145/1409944.1409948} based on energy detection \cite{20102413011425, 20124015492103} and feature detection \cite{20064610244009, 4289739}, respectively.
In \cite{20102413011425}, the authors studied energy detection by using discrete-time samples of signals.
In \cite{20124015492103}, the average performance of energy detection was derived by numerically integrating the detection threshold over the fading channels.
Besides, the authors of \cite{20064610244009} employed the peak detection in the high SNR regimes, and utilized the contour figure based pattern detection in the low SNR regimes.
In \cite{4289739}, the authors investigated cyclostationary feature of digital video broadcasting (DVB) signals, based on which a robust single-cycle detector was proposed to handle the uncertain noise issue. 
Despite the advancements, these methods suffer from three drawbacks, including the need for prior knowledge, high computational complexity, and low robustness against noise.

To address the above issues, another line of existing works \cite{8335463, 9547144, 9788039} employed DL for enhancing the signal detection performance.
The authors in \cite{8335463} implemented signal detection based on the time-frequency spectrum, in which CNNs are utilized for bounding box regression.
In \cite{9547144}, the authors investigated interference cancellation based on DL in faster-than-Nyquist communication systems. To reduce the continuous interference between adjacent signals, a long short-term memory (LSTM) algorithm-based recurrent neural network (RNN) was applied.
Moreover, in \cite{9788039}, the authors proposed a deep learning framework, which extracts features along the time axis at each frequency bin to predict the center frequency and the shape attributes of the signal. 

Despite the benefits in DL-based signal detection methods, these existing designs are not applicable for detecting the frequency and bandwidth of multiple coexisting signals. In contrast, our proposed detection method is able to not only detect the frequency and bandwidth of multiple signals, but also predict the modulation schemes.

\subsection{Automatic Modulation Classification}
Conventionally, the AMC is implemented via LB methods \cite{473837, 5351708, 6956852, 8362982} and FB methods \cite{2000415305186, 2004318286033, 6093991, 7728143}.
As for LB methods, a single-term approximation to the average log-likelihood-ratio was proposed in \cite{473837} to classify quadrature-modulated digital communication signals.
In \cite{5351708}, hybrid likelihood ratio test (HLRT) based and quasi HLRT based algorithms were investigated.
A blind modulation classifier for multiple-input multiple-output (MIMO) systems was proposed in \cite{6956852}, in which the channel matrix and noise variance were unavailable.
In \cite{8362982}, the authors investigated the LB-AMC for orthogonal frequency division multiplexing (OFDM) systems, in which two classifiers based on average likelihood ratio test and HLRT were proposed. In general, the LB methods are theoretically optimal in the Bayesian sense. However, they suffer from high computational complexity and require prior knowledge of both the signal and the channel.

On the other hand, the FB-based methods focus on feature extraction and classifier design.
The authors in \cite{2000415305186} introduced wavelet transform (WT) to extract the transient characteristics in a digital signal.
Homogeneous feature-vectors based on cyclic cumulants were proposed in \cite{2004318286033}, and the discrimination capability of the examined feature-vectors were verified.
Cyclostationarity-based features were utilized in \cite{6093991} to identify quadrature amplitude modulation (QAM) signals.
Order-statistics-based and reduced order-statistics-based AMC methods were proposed in \cite{7728143}, in which the approximate maximum likelihood and the back propagation neural networks classifier were introduced to the reduced order-statistics. 
Although FB methods have low computational complexity, they struggle to perform well under noise or multi-path channel fading.

In addition to LB and FB methods, DL-based methods have been recently adopted owing to huge amount of data available in modern communication networks. Specifically, the authors in \cite{7869126} introduced a spatial transformer model for AMC, which learns a localization network to blindly synchronize and normalize a radio signal without a prior knowing the signal structure.
Other neural networks, such as CNN, residual network, densely connected network, and convolutional long short-term deep neural network (DNN), were considered in \cite{8335483}.
Deep belief network and spiking neural network were introduced in \cite{9188007} to increase the classification accuracy for AMC in low SNR regimes.
In \cite{9220797}, a CNN and LSTM-based dual-stream structure was proposed, in which the features learned from two DL networks interact in pairs to increase the diversity of features.
Furthermore, a multitask learning-based DNN was proposed in \cite{9462447}, where three blocks, including CNN blocks, bidirectional gated recurrent unit (BiGRU) blocks, and a step attention fusion network (SAFN) block were interconnected to extract discriminative features.
In addition, a hierarchical classification head based convolutional gated DNN is proposed in \cite{9764618} by utilizing different layers’ output, which only utilizes the in-phase/quadrature cue and has a low computational cost.

\subsection{Real Signal Datasets}

Real signal datasets also do not target this complex signal environment. WiSig \cite{hanna2022wisig} treats detection as a data preprocessing step, which is already preprocessed for the purpose of WiFi source identification. The synchronization of the signal and location of signal existence have been preprocessed. This step is omitted and manually processed, without a separate algorithm to handle detection. There are also some related open-source research articles \cite{west2021wideband, vagollari2021joint}, but the corresponding datasets have not been publicly released. Currently, the signal processing industry is still mainly focused on baseband processing and modulation recognition, without considering the combination of signal detection and modulation recognition as a joint task.

In summary, existing methods and datasets mostly focused on the classification of one single clean signal, by ignoring the signal detection process before classification, in which the impact of potential detection bias is overlooked.
By contrast, this paper investigates the joint signal detection and AMC by explicitly considering the detection phase to avoid such limitations. 

\section{System Model}

\begin{figure}[htbp]
\begin{center}
\includegraphics[width=0.9\linewidth]{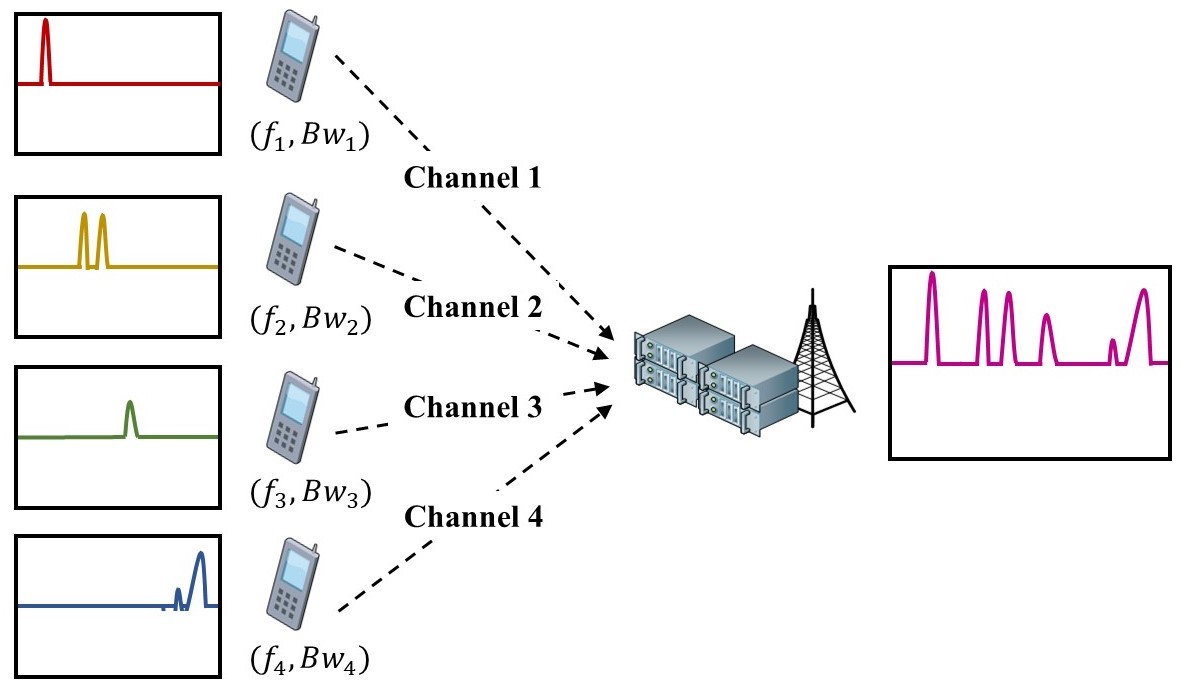}
\end{center}
   \caption{A typical example of complex signal environment.}
\label{fig:scene}
\end{figure}

As shown in Fig.~\ref{fig:scene}, we consider a typical complex communication environment with multiple signals, in which multiple transmitters may send radio signals over different frequency bands. Each transmitter first generates a digital transmit sequence, and then radiates it over the air after proper shaping. 
Next, a single-antenna receiver monitors the transmitted signals and employs blind modulation classification to analyze the modulation information without any prior knowledge.
The transmitted signal by the transmitter at a particular time $t$ is expressed as
\begin{equation}
\begin{aligned}
s_{i}(t)=g_i(t) \exp \left(-2 \pi f_i t\right),
\end{aligned}
\end{equation}
where $g_i(\cdot)$ represents the baseband modulation signal, which is shifted to the carrier frequency to meet specific transmission requirements, and $f_i$ represents the carrier frequency of the signal.

On the other hand, the received signal
is sampled by the receiver, which can be expressed as
\begin{equation}
\begin{aligned}
\begin{array}{r}
x[l]=\sum_{i=1}^n h_i  e^{j\left(2 \pi f_il + \theta_i\right)} s_i[l]+w[l], \\
l=1,2, \cdots, L,
\end{array}
\end{aligned}
\end{equation}
where $h_i$ represents the multipath channel coefficient that is a constant throughout the observation interval.
Let $\theta_i$ denote the phase offset, and $n$ denote the number of active transmitters.
The subscript $i$ represents each independent signal participating in the transmission process, and is eventually obtained at the receiver by superposition ($\sum_{i=1}^n$) to form the final spectrum.
The maximum number of samples is denoted as $L$, and $l$ represents the index of a specific sample. Moreover, $w(l)$ denotes the additive white Gaussian noise (AWGN).

The signal transmission is carried on an orthogonal basis with in-phase and quadrature (I/Q) components, respectively.
Thus, the baseband and passband I/Q signal is given by $s[n]=s_{\mathrm{I}}[n]+j s_{\mathrm{Q}}[n]$, and $s_{\mathrm{p}}[n]=\Re\left(s[n] \mathrm{e}^{j 2 \pi f_c n T_{\mathrm{s}}}\right)$, respectively, where $s_{\mathrm{I}}$ and $s_{\mathrm{Q}}$ represent the in-phase and quadrature components, $f_c$ is the center frequency, and $T_s$ is the sampling interval.

Once the information is captured, the raw complex signal is transformed into an I/Q sequence, which is given by
\begin{equation}
\begin{aligned}
x^{\rm{I/Q}}=\left(\begin{array}{l}
\mathrm{Re}[x(1), \ldots, x(L)] \\
\mathrm{Im}[x(1), \ldots, x(L)]
\end{array}\right),
\end{aligned}
\end{equation}
where $\mathrm{Re}[\cdot]$ and $\mathrm{Im}[\cdot]$ denote the real parts and the imaginary parts of the raw complex signal, respectively.

In the process of modulation, each transmitter encodes the information bits into the carrier signal by modifying its three primary characteristics including the amplitude, frequency, and phase. This alteration allows to represent analog signals in digital form. 

 In this paper, we aim to analyze a segment of a specific range of bands, detect a variable number of signals, and output independent modulation predictions by combining the characteristics of each signal. Specifically, the task is divided into three phases: data collection, data preprocessing, and task execution. The task is a two-stage sequential process, with the first detection sub-task, followed by the subsequent classification sub-task.

The goal of the first sub-task detection is to identify the center frequencies and bandwidths of existing signals for a supervised spectrum range. The detection is given by
\begin{equation}
\begin{aligned}
\{(c_i,w_i)\}_i^N=d(x[l], g),
\end{aligned}
\end{equation}
where $x[l]$ stands for the input signal frame, and $d(\cdot, \cdot)$ denotes the neural network structure for signal detection with $g$ standing for the weight vector of the neural network. Furthermore, let $N$ denote the number of detected outputs to be produced, consisting of different center frequency $c_i$ and different bandwidth $w_i$ pairs. 

When the detection sub-task is finished, the modulation classification of any detected signal $(c_i, w_i)$ can be made by
\begin{equation}
\begin{aligned}
H_i^k: \underset{1 \leq k \leq K}{\arg \max }=P\left(p_i^k \mid \boldsymbol{s}_i, c_i, w_i\right),
\end{aligned}
\end{equation}
where $k$ is the index of a certain modulation type and $K$ represents the total number of possible modulation schemes, $p_i^k$ denotes the probability of $i$-th singal being the $k$-th modulation type, and $\boldsymbol{s}_i = \{ s_i [1], s_i [2], \cdots, s_i [L] \}$ denotes the vector of baseband signal.

\section{Proposed Synthetic Dataset}

In this work, we generate a novel dataset, CRML23, for multiple-signal scenarios. In the following, we present the toolchain for generating compliant data entries. To improve the applicability of the dataset, we incorporate five widely utilized modulation schemes, including BPSK, QPSK, 8PSK, 16QAM, and 64QAM.
In particular, we define each individual item in the dataset as an ``entry", and each entry may contain multiple ``signals" to avoid confusion. This differs from the traditional datasets where each entry only includes one signal. In the subsequent sections, our proposed JDM will handle each entry as a single unit and separate different signals contained within each entry.

We utilize a recursive function to generate a complete entry with randomness. Within the main program and each recursive sub-program, we first determine whether the randomly selected bandwidth $W$ exceeds the specified upper limit $f_H$ or lower limit $f_L$. If $W>(f_H - f_L)$, which means that the signal to be generated falls outside the allowable bounds, it will not undergo further generation. Conversely, if the selected bandwidth falls within the permitted range, then we will verify whether the specific bandwidth remains unoccupied. Subsequently, we generate a baseband signal with a random modulation type within the range $(s_H, s_L)$ and accordingly generate the carrier frequency signal. Furthermore, a low-pass filter is utilized to prevent high-frequency leakage of the simulated baseband signal, thus avoiding interference with unoccupied bandwidth.
Following this, the function recursively handles the remaining unoccupied portions on both sides of the signal, namely $(f_L, s_L)$ and $(s_H, f_H)$. Via this iterative process, the program gradually fills the bandwidth within the specified range. The detailed dataset generation procedure is summarized in Algorithm~\ref{algorthm_dataset_generation}.

\input{tables/Algorthm}

Our dataset generation algorithm is capable of generating a diverse range of signal compositions. By introducing randomness in the generation of bandwidths within a predefined range and utilizing recursion, the algorithm can automatically generate signals. These signals are generated without any predetermined constraints. The upper bound and low bound on the number of signals are solely determined by the boundaries of the bandwidth range, allowing for unrestricted signal compositions within this range. Therefore, it is likely that some entries may not contain any signals, thus reflecting the diversity encountered in real-life scenarios. This variability is crucial for the development of analysis algorithms reliant on datasets, as the trained models must not only learn to handle multiple signals but also discern the cases where no signals are present.

\begin{figure}[htbp]
\begin{center}
   \includegraphics[width=\linewidth]{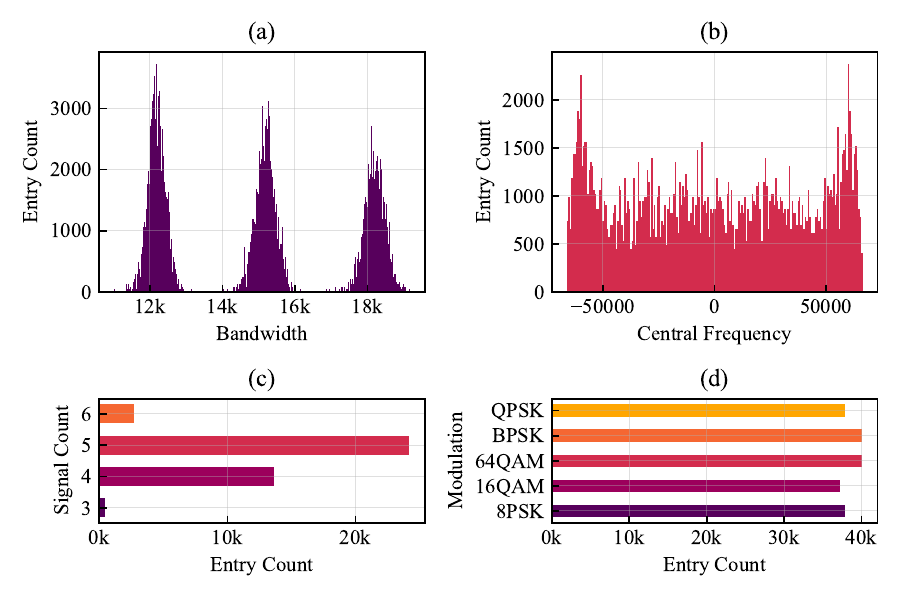}
\end{center}
   \caption{Subfigures (a) and (b) illustrate the total signal count per entry and the distribution of different modulation types within CRML23, respectively. Subfigures (c) and (d) depict the distribution of signal bandwidth and frequency, respectively.}
\label{fig:Distribution}
\end{figure}

As we can observe, CRML23 can offer rich features with randomness due to the randomly generated entries based on randomly selected channel models. In particular, each signal within the entry is independently generated with unique information, and the modulation type is also randomly selected. The distribution of bandwidth and center frequencies is illustrated in Fig. \ref{fig:Distribution}(a) and Fig. \ref{fig:Distribution}(b), respectively. The distribution of bandwidth exhibits three distinct clusters, resulting from the symbol rate specified in the signal generation algorithm. In the experiments, we will propose appropriate evaluation metrics that leverage this characteristic. Moreover, the frequency points exhibit a uniform distribution. In particular, the authors in \cite{zhang2022machine} have overcome the multiple-signal processing challenge by uniformly dividing the spectrum into multiple channels and generating a simulated dataset under the assumption that each transmitter occupies only one channel. However, this assumption introduces a restriction that the transmitter can only occupy pre-defined band. Moreover, the frequency points are limited within a few independent units, resulting in a clustered non-realistic distribution. 

As shown in Fig. \ref{fig:Distribution}(c), the number of signals per entry exhibits substantial randomness. This unpredictability arises due to the inherent uncertainty introduced by the recursive algorithm with random bandwidth allocation. The instances with an entry containing three or six signals, are relatively uncommon, while the entries with four to five signals are more frequently observed due to the imposed bandwidth ranges. Such generation patterns pose significant challenges for subsequent analysis. Additionally, as shown in Fig. \ref{fig:Distribution}(d), the occurrence frequencies of different modulation types in the generated signals are approximately equal, which is aligned with the manifestation of randomness within large-scale generation.

\input{tables/GenerationParameter}\

In the simulation, the channel fading is modeled as a stochastic factor. CRML23 comprises two fading channel models: Rayleigh fading and Rician fading. These models are differentiated by the presence or absence of line-of-sight (LoS) paths between the transmitter and the receiver. We employ these two models due to their widespread usage, while other channel models can be generated using the same methodology. The simulation parameters for the generation of CRML23 are summarized in Table \ref{tab:GenerationParameters}.

\section{Proposed Deep Learning-based Algorithm}

In this section, we elucidate the comprehensive architecture of the proposed JDM. In the following, we first present the overall structure of JDM, and then discuss the signal detection and the AMC modules, respectively.

\begin{figure*}[t]
\begin{center}

\includegraphics[width=\linewidth]{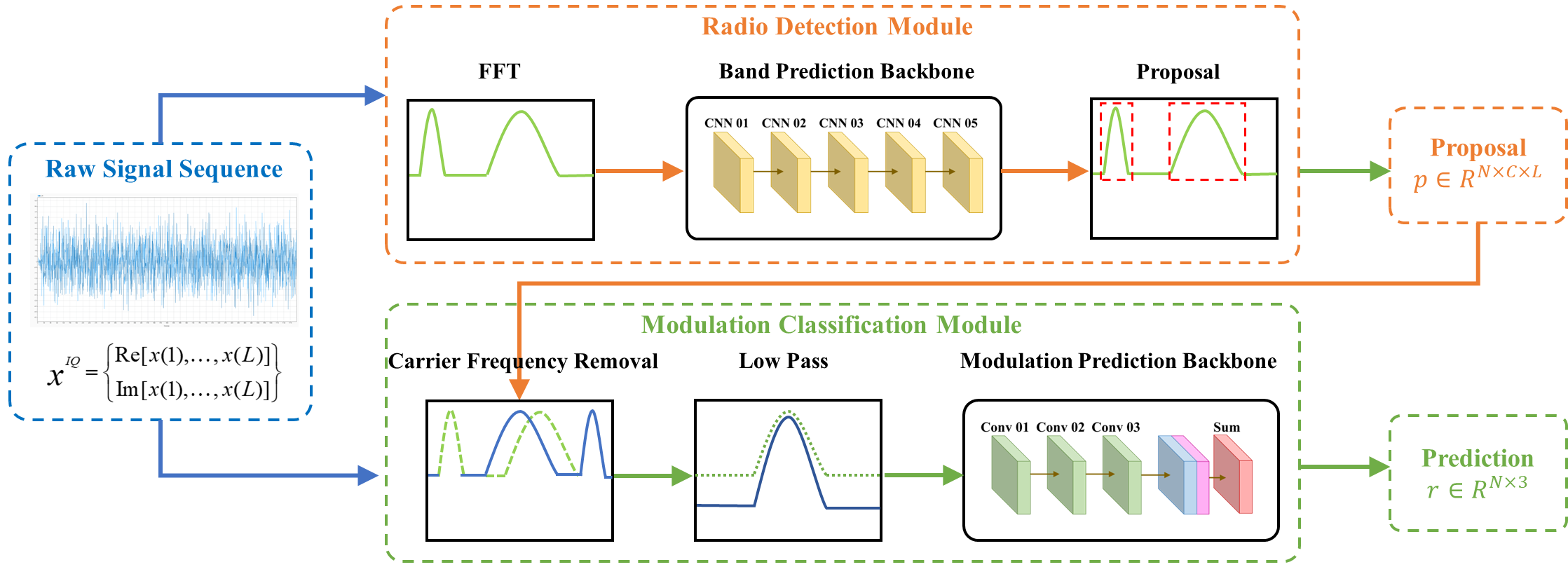}
\end{center}
   \caption{Joint framework for signal detection and automatic modulation classification (JDM).}
\label{fig:Framework}
\end{figure*}

\subsection{Overall Pipeline of JDM}\label{4-1}

Similar to the conventional synthetic signal data, our CRML23 datatset necessitates the assignment of corresponding labels to distinguish different modulation patterns. However, this distinction is a crucial prerequisite since misidentifying multiple signals as a single signal may lead to unpredictable consequences.
To address this issue, we propose a joint framework, as known as JDM, to address the tasks of complex signal detection and AMC at the same time. Note that our proposed JDM is decomposed into two distinct components, namely the signal detection module and the modulation classification module. Specifically, we adopt the concept of target detection to establish pathway between these
modules. In this way, we can exploit the relationships within the data to improve the accuracy in signal detection.

Upon inputting the original data stream as an I/Q sequence into the signal detection module, a fast fourier transform (FFT) is implemented to convert the time-domain signal into its frequency-domain representation. Then, a band prediction model is incorporated at the end of the signal detection module to generate predictions regarding the frequency location of the signal, referred to as “proposal". Following this, an AMC module is utilized to predict the modulation pattern of each filtered sequence of pure signals. These predictions are derived by considering the frequency and bandwidth characteristics of the different signals, culminating in the final output. The overall pipeline of JDM is depicted in Fig. \ref{fig:Framework}.

\subsection{Signal Detection Module}\label{4-2}

As an integral component of the signal processing flow, the signal detection module serves as the initial stage, which takes raw I/Q sequence data of the signal as input. Its primary function is to detect individual entries within the dataset, while simultaneously generating predictions for the center frequencies and bandwidths of multiple signals. These predictions aid the modulation classification module in making informed decisions. To enhance the accuracy of band prediction by operating in the frequency domain, an FFT operation is performed on the baseband signal in advance.  The output vector dimensions post-FFT align with the input I/Q sequences, maintaining a $2 \times L$ structure, where $2$ signifies the I and Q components and $L$ is the sequence length. 

To accomplish the task of signal detection, we propose a novel spectral model inspired by the YOLO model for image detection \cite{redmon2016you}. Bounding boxes are generated based on the I/Q sequence, representing a direct transformation of the predicted signal frequency and bandwidth. To simultaneously predict multiple bounding boxes with confidence scores, a CNN network is employed. The proposed spectral model is trained using comprehensive multiple-signal data, and its regression pattern mitigates performance degradation stemming from the complexity of the model's pipeline design. Furthermore, the utilization of global inference during training enables the model to capture the entire spectrum, thereby leveraging the complete context to unveil potential connections within the data. The utilization of candidate boxes offers the advantage of swift and comprehensive detection, which aligns with the dataset generation process. Based on such end-to-end training, our proposed model can achieve high accuracy and efficiency.

To handle an original I/Q signal sequence with batch size $N$, the network reconstructs several detection units. Among them, the detection unit closest to the center of the target signal is deemed to represent the signal. Within each detection unit, there are bounding boxes of varying sizes, corresponding to different possible bandwidths. The primary task of these bounding boxes is to predict the likelihood of a signal's presence within them. The confidence level is quantified using the Intersection over Union (IoU) metric, defined as
\begin{equation}
\begin{aligned}
\mathrm{IoU}_{\rm{pred}} = \frac{B_{\rm{gt}} \cap B_{\rm{pred}}}{B_{\rm{gt}} \cup B_{\rm{pred}}},
\end{aligned}
\end{equation}
where $B_{\rm{gt}}$ denotes the ground-truth box, representing the range occupied by the bandwidth of the ground-truth signal, and $B_{\rm{pred}}$ denotes the predicted box, indicating the range occupied by the predicted signal. For each bounding box, the model performs regression to determine a clustering center that corresponds to the range of the predicted bandwidth. If no object exists within the monitored range, the confidence level is 0.

In the training phase, each bounding box is designated to detect an individual signal, and objects are assigned to predictors based on the highest IoU scores. This specialized approach enables the model to learn the modulation pattern and size of the target signal, thereby improving the overall recall value. Moreover, the presence of signals within the detection module is assessed through the signal detection module and the modulation classification module, providing a signal detection mechanism that further enhances the recall value. The details of the mechanism will be discussed in Section \ref{4-3}.

\begin{figure}[htbp]
\begin{center}
   \includegraphics[width=\linewidth]{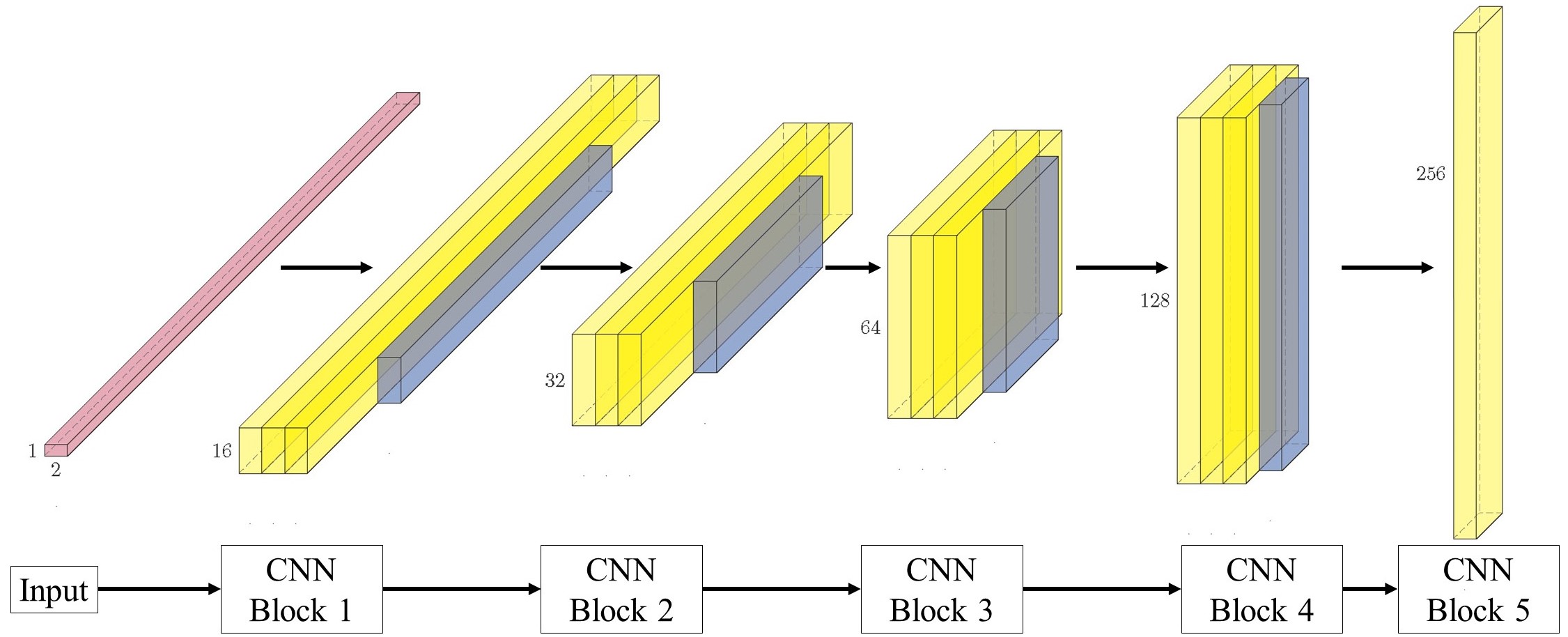}
\end{center}
   \caption{The structure details of the signal detection module.}
\label{fig:Signal Detection Module}
\end{figure}

To achieve signal detection, we employ a CNN architecture,
as depicted in Fig. \ref{fig:Signal Detection Module}, which is composed of five CNN blocks connected in a sequential manner.
Each block encompasses three convolutional layers, one rectified linear unit (ReLU) layer, and one batch normalization layer. The input data consists of preprocessed I/Q sequences denoted as ${x}_i^{\rm{I/Q}}$, where for any index $i$, ${x}_i^{\rm{I/Q}} \in {\mathbb{R}}^{N\times 2 \times 1200 \times 1}$. The dimensions include batch size, height, width, and channel. Multiple convolutional layers are employed to extract high-level semantic information from the signals, with the output of the last block utilized for prediction. ReLU introduces non-linearity, enabling the capture of complex features and mitigating overfitting issues associated with relatively simple data forms. Batch normalization plays a vital role in normalizing and linearly transforming data between layers, and decoupling the inter-layer dependencies. This normalization process ensures stable data ranges at the output of each layer, thereby enhancing model convergence and reducing sensitivity to network parameters. While some key parameters vary across layers, padding is uniformly set to $(0,0)$. The final output ${x}_i^{\rm{CNN}} \in {\mathbb{R}}^{N\times 1 \times 144 \times 256}$ represents the feature maps consisting of 256 channels.

The output of the network is denoted by ${\mathbf{p}}_i \in \mathbb{R}^{N\times C \times L}$, which is referred to as proposal, represented as a matrix $[[f_c^1,B^1], \cdots, [f_c^N,B^N]]$. This output contains predictions regarding the location and bandwidth of each signal in the spectrum. Here, the dimension $L$ is similar to the feature map size in computer vision applications. However, unlike the multidimensional scenarios in computer vision, our signal data is one-dimensional. Consequently, $L$ denotes the number of detection units within our framework, with each unit acting as the smallest divisible scale that accommodates an anchor. $C$ denotes the number of available anchor, which is set to be 3 in our case. This implies that each bounding box corresponds to three predictions: $B$ (predicted bandwidth), $f_c$ (central frequency location), and confidence level. $N$ represents the number of the data input, which could be seen as the batch size in deep learning, recording the number of entries. The predicted bandwidth and central frequency jointly determine the upper bound and lower bound of the spectrum occupied by the signal, defined as $(f_c-\frac{1}{2}B, f_c+\frac{1}{2}B)$. The confidence level indicates the IoU between the predicted box and the ground-truth box. When the confidence level surpasses a specified threshold, the bounding box is considered a positive sample; otherwise, it is regarded as a negative sample.

\subsection{Modulation Classification Module}\label{4-3}

The input of the modulation classification module includes two parts: the raw signal data in the form of I/Q sequences and the proposals generated by the signal detection module. 
Leveraging the information provided by the proposals, the original data is processed to obtain a pure spectrum that solely contains a single signal with its individual modulation type. The carrier frequency is eliminated from the original signal by utilizing the carrier information predicted in the proposal. Furthermore, prior to modulation analysis, a low-pass filter is applied to mitigate the noise from the signal. The primary purpose of employing a low-pass filter is to filter out high-frequency noise that may be present in the signal, thereby enhancing signal quality and clarity. The passband frequency is carefully chosen to ensure that essential components of the signal are preserved, while effectively reducing components beyond this threshold that are likely to be noise artifacts. In particular, a finite impulse response (FIR) filter is utilized to realize this function. The resulting filtered signal, denoted by $\mathbf{x}_i^{\mathrm{I/Q}}$, serves as the final output of the preprocessing stage and is fed into the backbone network. 

\begin{figure}[htbp]
\begin{center}
\includegraphics[width=0.9\linewidth]{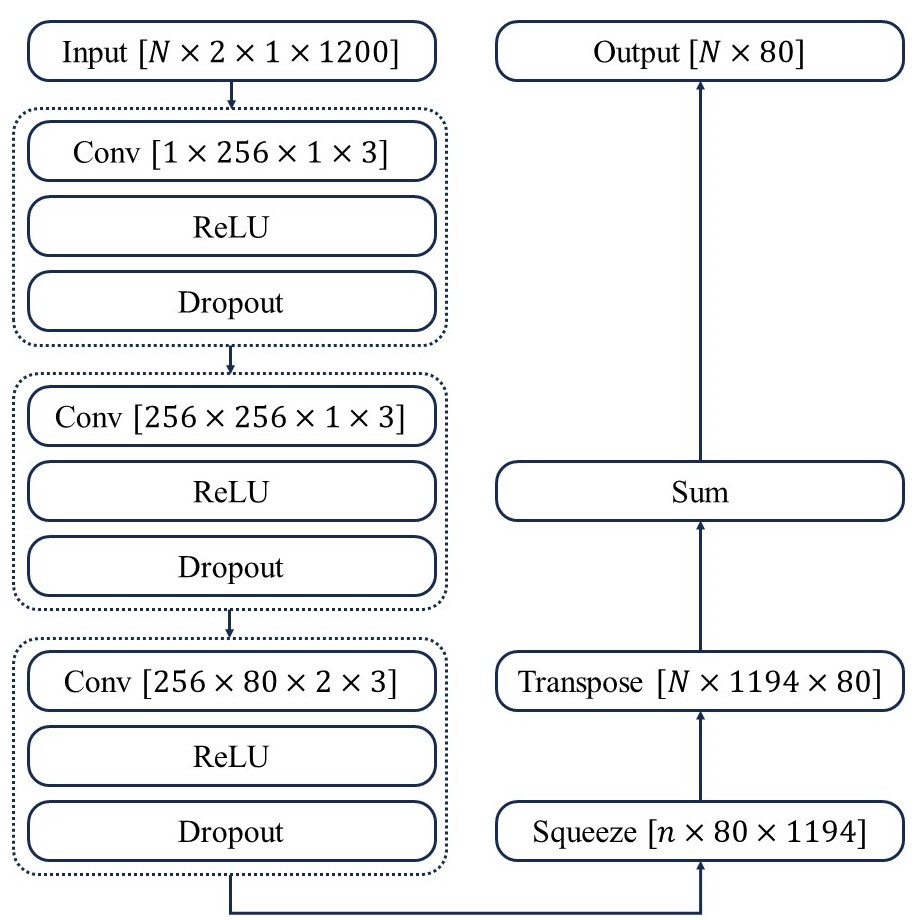}
\end{center}
   \caption{The structure of neural network in modulation classification module.}
\label{fig:Modulation Classification Module}
\end{figure}

As illustrated in Fig. \ref{fig:Modulation Classification Module}, the CNN block consists of three sets of CNN layers. Each set comprises three parts: a convolutional layer, a ReLU layer, and a dropout layer. The input data is a preprocessed I/Q sequence denoted as $\mathbf{x}_i^{\mathrm{I/Q}}$. For any index $i$ in the sequence, we have $\mathbf{x}_i^{\mathrm{I/Q}} \in \mathbb{R}^{N\times 2 \times 1200 \times 1}$, where each dimension represents the shape of the filter bank in the first convolutional layer. Specifically, $\mathbf{W}_1^{\mathrm{CNN}} \in \mathbb{R}^{1 \times 256 \times 1 \times 3}$ denotes the filter bank's height, width, channel, and the number of filters, respectively. The primary function of this layer is to map the low-dimensional data to a high-dimensional space. Subsequently, two additional convolutional layers, $\mathbf{W}_2^{\mathrm{CNN}} \in \mathbb{R}^{256 \times 256 \times 1 \times 3}$ and $\mathbf{W}_3^{\mathrm{CNN}} \in \mathbb{R}^{256 \times 80 \times 2 \times 3}$, are applied. The final output data $\mathbf{x}_i^{\mathrm{CNN}} \in \mathbb{R}^{N\times 1 \times 80 \times 1194}$ represents a feature map with 80 channels. The inclusion of ReLU introduces nonlinearity, capturing complex features and mitigating overfitting issues caused by the relatively simple data form. Dropout is similarly employed within our skeleton network to alleviate the overfitting problem arising from the network depth. For parameters that remain consistent across layers, padding is set to $(0,0)$, the stride is set to $(2,2)$, and the dropout rate is set to 0.5.

The output of the CNN, denoted by $\mathbf{x}_i^{\mathrm{CNN}}$, undergoes two reshape layers to match the input requirements of the Sum layer. A squeeze layer is utilized to eliminate dimensions of size 1 from the input, reducing the data dimension without compromising information content. The output is denoted by $\mathbf{x}_i^{\mathrm{sq}} \in \mathbb{R}^{N \times 80 \times 1194}$. Subsequently, to ensure that the Sum layer operates on channel content rather than data width, we transpose the data by swapping two sizes. This reshapes the data to $\mathbf{x}_i^{\mathrm{trans}} \in \mathbb{R}^{N \times 1194 \times 80}$. The transposed data is then fed into the Sum layer, which performs summation along the second axis, resulting in $\mathbf{x}_i^{\mathrm{output}} \in \mathbb{R}^{N \times 80}$.

\section{Experiment Results}

In this section, we first introduce the evaluation metrics to assess the prediction performance of our proposed JDM. Then we conduct a series of experiments to show the impact of several key factors on system performance, including Rayleigh and Rician channels, Doppler frequency shift, K-factor, SNR, clock offset, and the influence of different modulation schemes on the model's prediction accuracy.
Our model, trained on a varied dataset, is designed for robust performance across different signal conditions without segregating by channel types like Rayleigh or Rician during the training phase. This ensures versatility in handling various environment variables. The model's adaptability is further enhanced by exclusively testing on specific channel conditions, offering a focused assessment of its real-world applicability. 

To provide further information, we also evaluate a blend of traditional methods tailored for signal processing challenges. Specifically, we employ Matched Filter (MF) and Threshold Judgment (TH) as signal detection techniques, alongside Decision Tree (DT) and Support Vector Machine (SVM) for modulation classification. This combination allows us to assess the efficacy between these methodologies with our framework.

For the implementation, we utilize the PyTorch \cite{paszke2019pytorch}, a DL platform running on a Linux server, equipped with Nvidia RTX3090TI GPU, 120GB RAM, and an Intel Xeon Silver 4214 CPU. During the training phase, we employ the Adam optimizer for detection module, and AdamW optimizer for classification module, due to the advantage of Adam(W)'s stability and effectiveness compared to SGD. The batch size is set to be 12 and 32 for detection and classification, respectively, and the learning rate is set to 0.001, with a weight decay of 0.00005 for AdamW. We train the model over a total of 90 epochs, including 30 epochs for detection, and 60 epochs for classification, which are chosen to achieve the maximum classification accuracy.

\subsection{Evaluation Methods}
In machine learning, the relationship between prediction results and ground-truth labels can be classified into four cases: true positive (TP), false positive (FP), false negative (FN), and true negative (TN). Among them, two metrics can further evaluate the training performance of the overall sample, namely Precision, and Recall, which are expressed by
$\text{Precision}=\frac{T P}{T P+F P}$, $\text{Recall}=\frac{T P}{T P+F N}$, respectively.

Based on the above two metrics, we create a precision-recall curve (PR curve) to demonstrate the internal link between the two and to measure the performance of the signal detection model. In order to combine the two metrics and assess the performance of the model with a single key metric, we introduce key performance evaluation metrics in the field of target detection, namely mean Average Precision (mAP).

We take the decrease of the maximum precision value as a trigger condition for sampling, and record each recall sample point that needs to be involved in the calculation. Geometrically, the calculation process can be defined as the area of the smoothed PR curve with an X-axis envelope, i.e., area under curve (AUC). To this end, the mAP is expressed as
\begin{equation}
\begin{aligned}
{\rm{m A P}} &=\frac{\sum_{i=1}^K {\rm{A P}}_i}{K}
{\rm{A P}}, \\
&=\sum_{i=1}^{n-1}\left(r_{i+1}-r_i\right) p_{\rm{interp } }\left(r_{i+1}\right),
\end{aligned}
\end{equation}
where $K$ is the total number of categories to be classified, which is expressed as the number of modulation patterns in this paper.

Inspired by the COCO dataset, we extend our evaluation system by adding more AP calculation methods. The evaluation metrics can be further divided into two categories, IoU and pixel area. Different IoU thresholds are utilized as a filtering condition, and ${\rm{A P}}^{.50}$ and ${\rm{A P}}^{.75}$ present the AP measurements for IoU thresholds of 0.5 and 0.75, respectively. In addition, by taking into account the size of the target box, let ${\rm{A P}^{small}}$, ${\rm{A P}^{medium}}$ and ${\rm{A P}^{large}}$ correspond to the AP values for different sizes of bounding boxes with pixel areas of $(0^2,32^2)$, $(32^2, 96^2)$, and $(96^2, \infty^2)$.
As a transformation, we manually establish two thresholds based on the distribution of bandwidth in Fig. \ref{fig:Distribution}(a), which correspond to three types of bounding boxes: $(0, 110)$, $(110, 130)$, and $(130, 150)$. In this context, we choose the number of samples as the unit for ease of practical computation, rather than the signal bandwidth. These thresholds serve as the evaluation criteria for ${\rm{A P}^{small}}$, ${\rm{A P}^{medium}}$, and ${\rm{A P}^{large}}$, respectively.

We also incorporate another metric, namely average recall (AR), to assess the predictive performance of the model. Drawing from the definition employed in the COCO dataset, AR is computed by averaging the recall values across various IoU thresholds spanning from 0.5 to 1.0. These thresholds are derived by considering twice the area under the curve obtained from plotting recall against IoU. In this paper, however, we categorize signals based on their quantity rather than class. By analyzing the distribution of the actual number of signals contained in each entry of CRML23, we find that the top three signal quantities are 4, 5, and 6, accounting for $33.3\%$, $59.0\%$, and $6.0\%$, respectively. Hence, we select these three parameters as the benchmark for AR calculation and conduct subsequent analysis in the experimental phase.

Similarly, we also examine the impact of signal size on the AR evaluation metric, taking into account different scales. The definition of these sizes aligns with the content discussed earlier regarding mAP.

\subsection{Evaluation on Detection Module}

\begin{figure}[htbp]
\begin{center}
   \includegraphics[width=\linewidth]{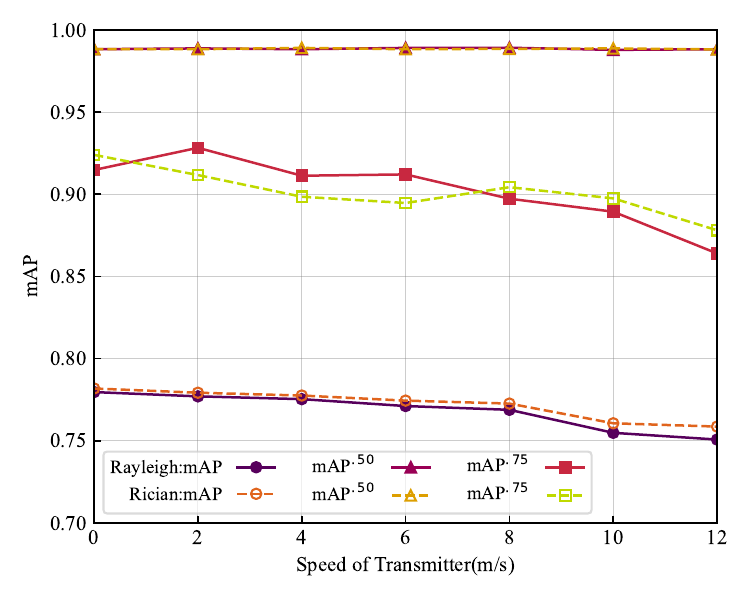}
\end{center}
   \caption{Effect of object velocity on \textbf{detection module}'s mAP under Rayleigh and Rician channel. Solid curves represent the Rayleigh channel, while hollow curves represent the Rician channel.}
\label{fig:Graph1}
\end{figure}

Through experimental simulations of different velocities of relative motion, we observe the impact of Doppler frequency shift on signal transmission under Rayleigh and Rician channel conditions, while the K-factor was fixed at 4. Results are shown in Fig.\ref{fig:Graph1}. For both Rayleigh and Rician channels, as the Doppler effect strengthens (i.e., an increase in the velocity of the transmitting end), the mAP shows a gradual decline. This indicates a reduced ability of the model to accurately detect targets due to the blurring and deformation of target features caused by Doppler frequency shift. Comparing the performance between the two channel types, it is observed that Rayleigh channels outperform Rician channels in terms of accuracy. This may be attributed to the presence of a LoS component in Rayleigh channels, which enhances the reliability and robustness of the signals. Further analysis of the relationships among mAPs with different IoU reveals that for stricter IoU thresholds, the accuracy of the model further decreases.

\begin{figure}[htbp]
\begin{center}
   \includegraphics[width=\linewidth]{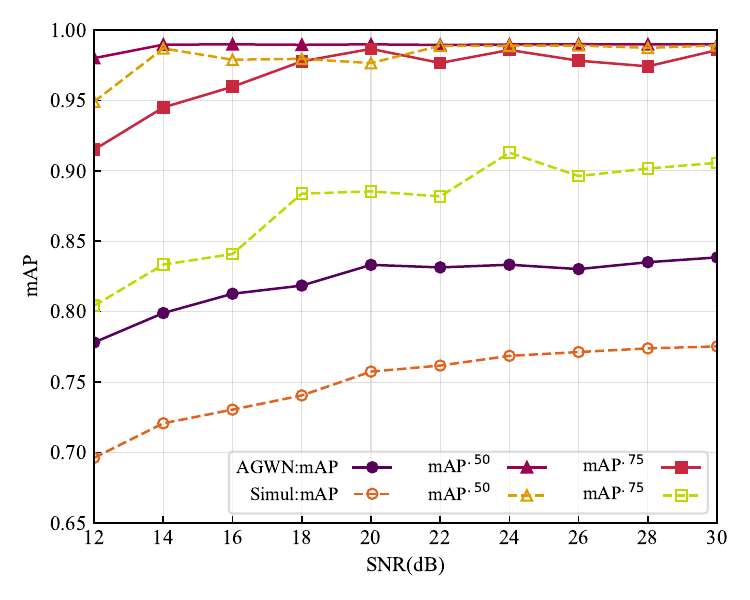}
\end{center}
   \caption{Effect of SNR on \textbf{detection module}'s results under \textbf{AGWN and simulated settings}. 
Solid shapes represent the AGWN setting, while hollow shapes represent the simulated setting.}
\label{fig:Graph2}
\end{figure}

Subsequently, we analyze the accuracy performance in an AWGN environment in Fig.\ref{fig:Graph2}. In this environment, the velocity of the transmitting end is set to zero to eliminate the impact of velocity variations on the results. The progressive increase in SNR leads to a significant improvement in accuracy, as the reduced interference of noise on the signal makes the target features clearer and more distinguishable. We find that there exists an upper limit of accuracy convergence, where further increasing the SNR has little effect on improving detection accuracy. 

We also introduce random generation of parameters to simulate the diversity and uncertainty present in real-world scenarios. These factors include channel effects, data characteristics, object velocities, and K-factor. The setting composed of these factors is referred to as the ``simulated setting," which is also illustrated in Fig. \ref{fig:Graph2}. We observe a general decrease of approximately 10 percent in accuracy compared to the ideal AWGN environment. This can be attributed to the introduction of additional random factors, which further increase the difficulty for the model to learn the underlying patterns in the data. In more realistic simulations, the task of target detection at higher confidence thresholds becomes more challenging. 

\begin{figure}[htbp]
\begin{center}
   \includegraphics[width=\linewidth]{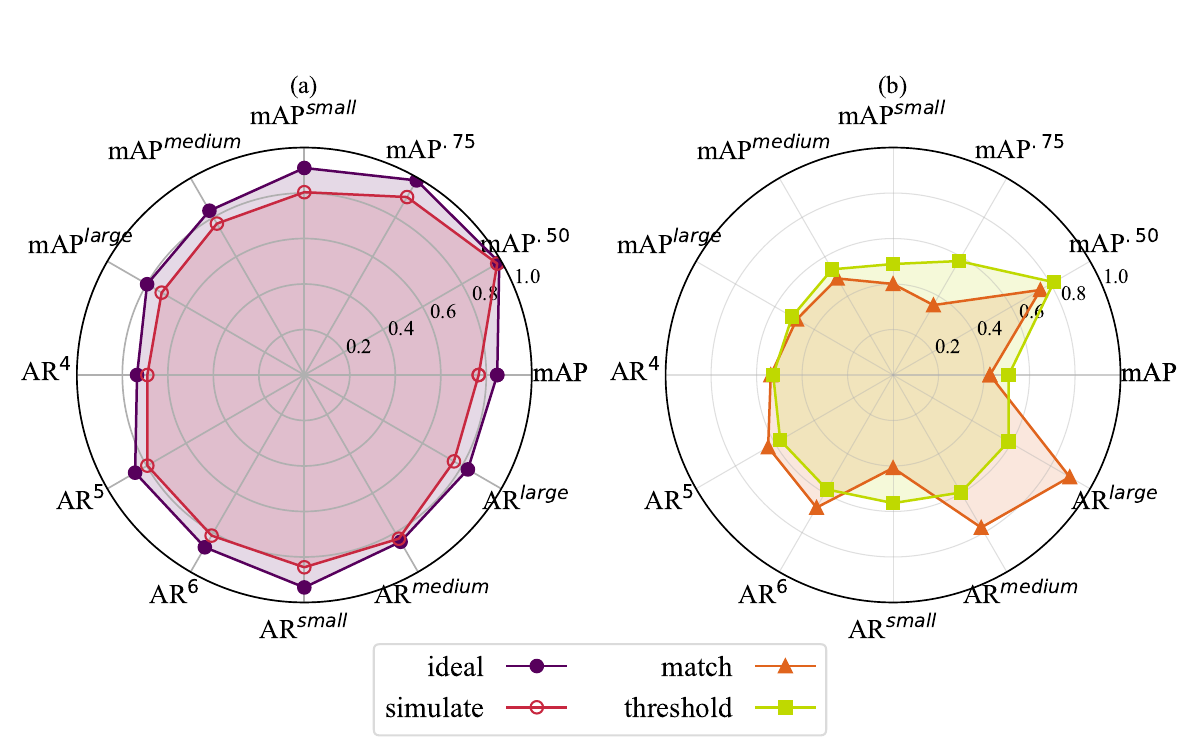}
\end{center}
   \caption{The evaluation metric. (a) presents the performance of the \textbf{detection module}, while (b) shows the performance of conventional methods.}
\label{fig:Graph3}
\end{figure}

In Fig.\ref{fig:Graph3}, we introduce the SNR as an additional random factor to test the accuracy of the detection model. We compare the performance in two signal environments: the pure signal environment without any random factors involved, referred to as the ``ideal setting," and the realistic signal environment with various interference taken into account, referred to as the ``simulated setting." Furthermore, we test two typical methods in traditional signal detection, namely matched filter detection and energy detection, using our CRML23 under the same signal conditions. The experimental results demonstrate the superiority of our proposed DL approach based on object detection principles. Particularly, concerning the mAP, which is sensitive to accuracy, our method exhibit compelling performance, firmly establishing its efficacy.

We observe small difference in detection accuracy between ideal conditions and simulated environments. This indicates that our model exhibits good robustness when facing real communication environments. Further analysis revealed that the detection accuracy for smaller-sized (narrow bandwidth) signals is higher, while the detection accuracy for larger-sized (wide bandwidth) signals is relatively lower. This suggests that narrower bandwidth signals have more concise features in the frequency domain, allowing the model to more accurately identify and locate targets. 

Although the mAP metric noticeably decreases in practical scenarios, the decrease in the AR metric is relatively smaller. This presents that AR is less sensitive to data fluctuations compared to mAP. We also notice a significant decrease in accuracy for the AR@4 metric. The reason is that AR@4 does not cover a substantial number of entries with five signals present in CRML23, resulting in lower accuracy. The difference between AR@5 and AR@6 is relatively small because such instances are relatively rare, leading to a lower probability for the model to learn this feature and providing a smaller gain in accuracy.

\subsection{Evaluation on Classification Module}

\begin{figure}[htbp]
\begin{center}
   \includegraphics[width=\linewidth]{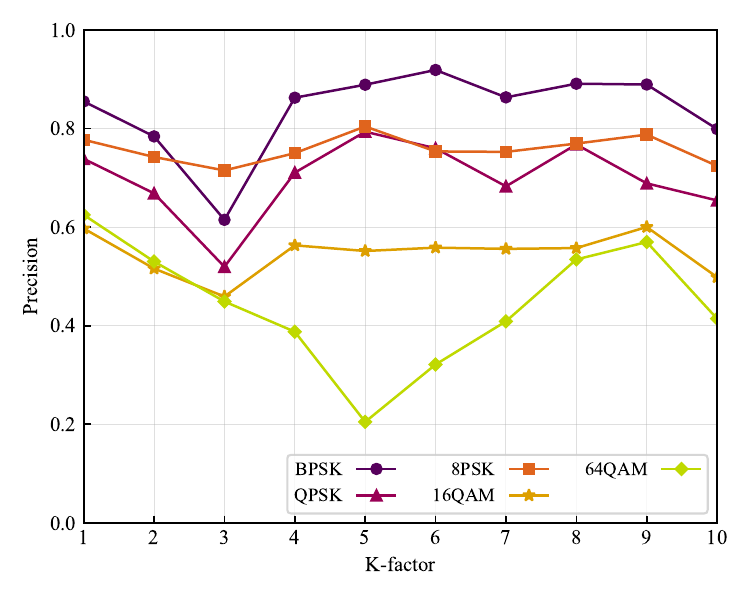}
\end{center}
   \caption{Effect of K-factor on \textbf{classification module}'s precision under Rayleigh channel.}
\label{fig:Graph4}
\end{figure}

The K-factor is a parameter utilized to measure the power difference between the LoS path and the reflected paths in a signal's multipath propagation environment. We conduct experiments to measure and analyze the impact of different K-factor values on signal transmission performance. In Fig. \ref{fig:Graph4}, we observe that different K-factor values had inconsistent effects on model accuracy, and no clear trend could be observed. This suggests that the properties of the Rician channel are not significantly influenced by the K-factor, or the model is not sensitive to performance variations with different K-factor values. The order of the modulation scheme still remains the primary factor affecting model accuracy, with higher complexity in modulation schemes resulting in lower recognition accuracy. However, we notice that for certain specific K-factor values, each modulation scheme exhibits similar tendencies. This may indicate that under specific channel conditions, the impact of modulation schemes on model performance is influenced by the K-factor.

\begin{figure}[htbp]
\begin{center}
   \includegraphics[width=\linewidth]{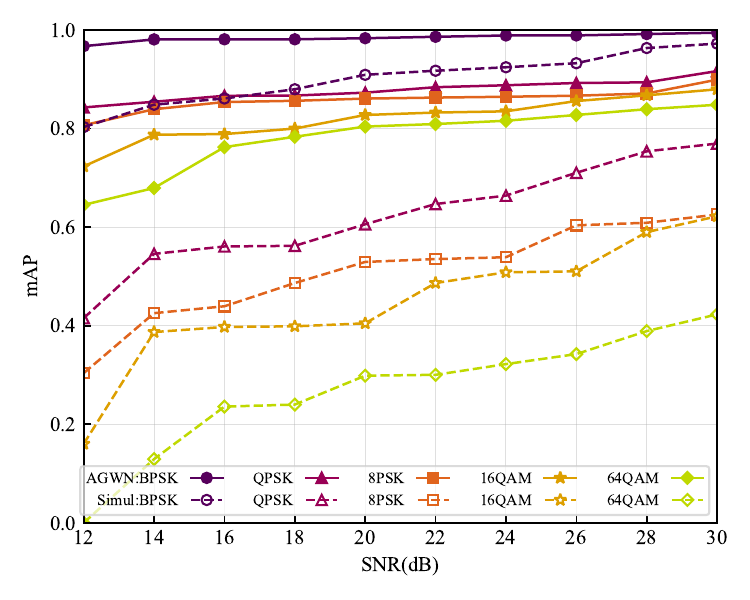}
\end{center}
   \caption{Effect of SNR on \textbf{classification module}'s results under \textbf{AGWN and simulated settings}. 
Solid shapes represent the AGWN setting, while hollow shapes represent the simulated setting.}
\label{fig:Graph5}
\end{figure}

Fig.\ref{fig:Graph5} demonstrates a significant improvement in prediction accuracy for the AMC model across all modulation schemes as the SNR increases. At higher SNR levels, the BPSK modulation scheme reaches a prediction accuracy close to 1.0 earlier, while the prediction accuracy of the other modulation schemes becomes similar. At this point, the influence of the modulation scheme's complexity on prediction accuracy is diminished. The accuracy of the model, which serves as an important metric for evaluating overall performance, also shows a corresponding improvement trend. However, it is important to note that there is a convergence upper limit in the experimental results. This may be attributed to the inherent characteristics of the modulation schemes and limitations imposed by the channel conditions. In the realistic environment, the degradation is more severe compared to the AWGN environment, especially under low SNR conditions. Additionally, higher-order modulation schemes experience greater degradation in the realistic environment. The increased complexity of the environment introduces more interference and channel fading to the transmission of higher-order modulation schemes, making it difficult for the model to learn patterns and effectively predict accuracy.

\subsection{Evaluation on JDM}

\begin{figure}[htbp]
\begin{center}
   \includegraphics[width=\linewidth]{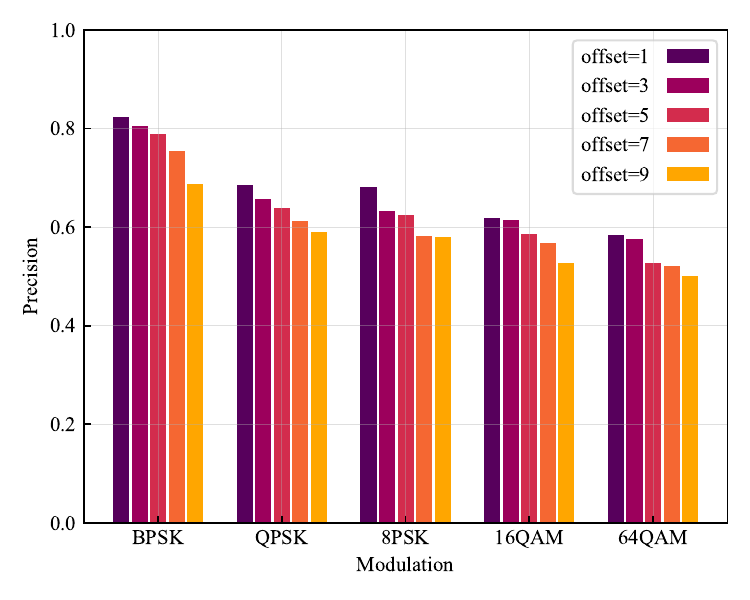}
\end{center}
   \caption{Effect of clock offset on the \textbf{proposed JDM}'s precision.}
\label{fig:Graph6}
\end{figure}

Clock offset refers to the difference in clock values between different devices in a communication system. In Fig.\ref{fig:Graph6}, we consider the clock offset under different modulation modes and evaluate its impact on the system performance by measuring the comprehensive detection accuracy. The results indicate that as the complexity of the modulation mode increases, the prediction accuracy generally decreases. This could be attributed to the mismatch between signal sampling time and synchronization caused by the clock offset, which subsequently affects the accuracy of target detection and localization. 

\begin{figure}[htbp]
\begin{center}
   \includegraphics[width=\linewidth]{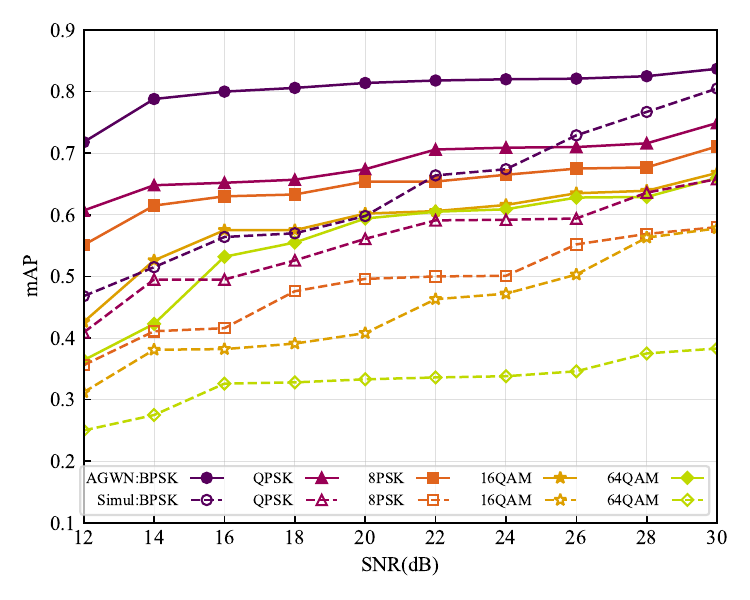}
\end{center}
   \caption{Effect of SNR on the \textbf{proposed JDM}'s precision under \textbf{AGWN and simulated settings}. Solid shapes represent the AGWN setting, while hollow shapes represent the simulated setting.}
\label{fig:Graph7}
\end{figure}

In Fig.\ref{fig:Graph7}, we investigate the impact of SNR on the detection accuracy of the proposed JDM. As the SNR increases, the received signals become clearer, enabling the target detection module and the modulation classification module to collaborate more effectively in accurately identifying and locating targets. Compared to the accuracy in the modulation classification module, we observe a general decrease in the framework accuracy by 20-30 percent. This can be attributed to the fact that the detection module focuses on enhancing target detection and localization performance, while the AMC accuracy tends to prioritize classification accuracy and is less sensitive to the target's position and scale information. Therefore, we speculate that the classification process could be a bottleneck factor that lowers the framework accuracy, rather than solely the influence of modulation mode selection. 

Meanwhile, a comparison of the JDM accuracy in a simulated environment is presented. Compared to AWGN, the framework accuracy generally decreases by 10 to 15 percent. In contrast to the individual output of the modulation classification module, the JDM accuracy exhibits a smaller variance in modulation mode accuracy at low SNRs. This indicates that the classification module struggles to accurately predict complex modulation modes in low SNR environments, resulting in smaller performance differences among the modulation modes. The comprehensive framework addresses this limitation and improves the fairness of performance by mitigating this drawback. However, as the SNR increases, the improvement in framework accuracy becomes relatively small. This is limited by the combined effects of the two modules, which aligns with the convergence limit.

\begin{figure}[htbp]
\begin{center}
   \includegraphics[width=\linewidth]{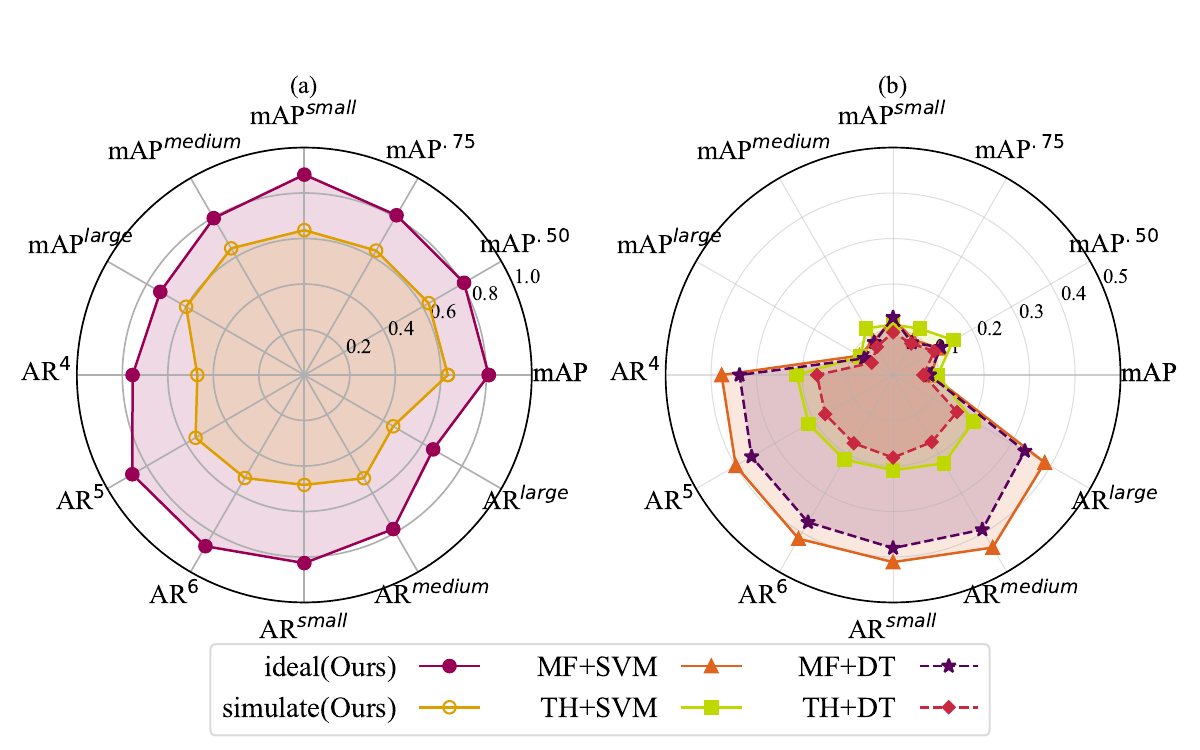}
\end{center}
   \caption{The evaluation metric. (a) shows the performance of \textbf{the proposed JDM}, while (b) shows the performance of conventional methods. Coordinates in (b) are enlarged for detailed observation.}
\label{fig:Graph8}
\end{figure}

Finally, when incorporating SNR as a random factor in the experiments, Fig. \ref{fig:Graph8}(a) illustrates the remarkable performance of the model under an ideal and simulated environment. In terms of AR, the model demonstrates high robustness across all detection scales. However, in the simulated environment, the model's object detection capability faces significant challenges due to presence of random interference and noise. Additionally, based on the data in the table, we also observe that in terms of size, small-sized objects exhibit higher prediction accuracy.

In comparison, experiments on the combination of traditional methods in Fig. \ref{fig:Graph8}(b). By incorporating Decision Tree and Support Vector Machine modulation classification approach, combined with Matched Filter and Threshold judgment methods in detection, the traditional methods exhibited lower accuracy, due to the following reasons. Firstly, the proposed dataset presents higher complexity due to the coexistence of multiple signals and high randomness. Secondly, without considering the impact of the detection task on the classification task, errors would accumulate and ultimately manifest in the results. Lastly, the limitations of traditional methods became apparent, as similar methods were unable to detect bandwidth. Their detection algorithms fundamentally rely on classification, avoiding some crucial issues present in real-world scenarios.

\section{Concluding Remarks}

In this paper, we introduced a simulated dataset, CRML23, generated from real signal environments for joint signal detection and modulation classification. The proposed dataset specifies a specific frequency band and includes a large number of signals, in which the parameters were randomly generated to achieve the highest level of realism. The generated dataset has various characteristics of the signals in each entry. 
Moreover, we proposed a novel joint framework, JDM, which performs the tasks of detection and classification. Two modules inside Coordinate with each other and pass information through the proposal. We demonstrated the effectiveness of CRML23 and discussed the impact on performance with a wide variety of parameters. Furthermore, we evaluated the performance gains achieved by different parameters and subnetworks within the framework.
The experimental results demonstrated that CRML23 and JDM achieved significant performances in terms of signal detection and modulation classification on raw data.

There are various interesting topics worth further pursuing in the future, which are discussed in the following. A notable limitation pertains to the performance of our algorithm in low signal-to-noise ratio (SNR) environments. This deficiency is primarily attributed to the classification module not being tailored explicitly for such conditions and the potential amplification of bias during the proposal transmission between the two modules.

Future efforts will concentrate on enhancing the classification module and the pathway between the two modules. In recognition of the current limitations of our modulation classification module, a key area for future development is to enhance its adaptability, particularly when the input shifts from baseband signals to processed, noisy, and complex signals. Our module currently lacks a design that adapts to this transition effectively. Future efforts will thus focus on optimizing the module algorithms to better handle such complex inputs.

Additionally, we will address the bottleneck issues that arise from the independent operation of the signal detection and modulation classification modules within our framework, which becomes evident in low signal-to-noise ratio environments. Developing joint algorithms and improving the communication and coordination between these modules will be pivotal in creating a more cohesive and efficient unified system.

\bibliographystyle{ieeetr}
\bibliography{ref.bib}





\end{document}

%% file: tables/Algorthm.tex
\renewcommand{\algorithmicrequire}{\textbf{Input:}}
\renewcommand{\algorithmicensure}{\textbf{Output:}}

\begin{algorithm}[htbp] \small
\caption{Dataset Generation Algorithm}
    \begin{algorithmic}[1]
        \Require
          $l_f$: last lower bound;
          $h_f$: last upper bound;
          $y$: last entry
        \Ensure 
            $y'$: new entry
        \State Randomly generate modulation, channel, noise, central frequency, and bandwidth $W$.
        \If {$W$ \textgreater ($f_H$ - $f_L$)}
            \State \Call{exit}{}
        \Else
            \State Generate a new signal $u$ with above parameters.
            \State Calculate a effective range $(R_H, R_L)$ for signal generation in $(f_L, f_H)$.
            \State Add signal $u$ in $(R_H, R_L)$.
            \State Lowpass operation, set carrier frequency and channel, update entry $y$.
            \State Set $s_L$, $s_H$ with $u$'s lower and upper bound.
            \State \Call{Algorithm 1}{$y$, $f_L$, $s_L$}
            \State \Call{Algorithm 1}{$y$, $s_H$, $f_H$}
        \EndIf
    \end{algorithmic}
\label{algorthm_dataset_generation}
\end{algorithm}

%% file: tables/GenerationParameter.tex
\begin{table}[hbpt]
  \centering
  \caption{{{Simulation Parameters}}}
  \renewcommand\arraystretch{1.25}
  \begin{tabular}{|m{4cm}<{\raggedright}| m{3cm}<{\raggedright}|}
    \hline
    \textbf{Abbreviations} & \textbf{Distribution} \\
    \hline
    Sample rate                      
    & $1.5\times 10^{5}$ Hz              \\
    \hline
    SNR                             
    & [12:30:2] dB                         \\
    \hline
    Path delays                       
    & $[0, 1.8, 3.4]\times 10^{-7}$       \\
    \hline
    Average path gains               
    & 0, -2, -10                \\
    \hline
    Kfactor                       
    & [1:10:1]                         \\
    \hline
    Maximum doppler shift           
    & 4                         \\
    \hline
    Maximum clock offset            
    & 5                         \\
    \hline
    Center frequency                 
    & [0:36:6] Hz                     \\
    \hline
    Channel model                 
    & Rician / Rayleigh           \\
    \hline
\end{tabular}
\label{tab:GenerationParameters}
\end{table}